\documentclass{article}

\usepackage{arxiv}

\usepackage[utf8]{inputenc} 
\usepackage[T1]{fontenc}    
\usepackage{hyperref}       
\usepackage{url}            
\usepackage{booktabs}       
\usepackage{amsfonts}       
\usepackage{nicefrac}       
\usepackage{microtype}      
\usepackage{cleveref}       
\usepackage{graphicx}
\usepackage{doi}
\usepackage{etaremune}
\usepackage{textcomp}

\title{How does a spontaneously speaking conversational agent affect user behavior?}

\date{April 30, 2022}

\author{ Takahisa Iizuka \\
	Graduate School of Regional Development and Creativity\\
	Utsunomiya University\\
	7-1-2 Yoto, Utsunomiya, 321-8585 Japan \\
	\texttt{iizuka@speech-lab.org} \\
	\And
	\href{https://orcid.org/0000-0002-2883-1143}{\includegraphics[scale=0.06]{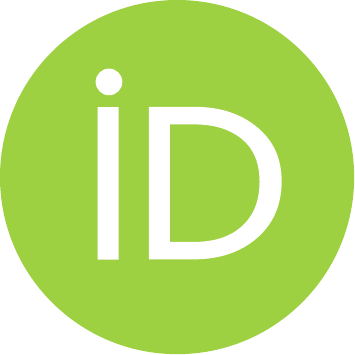}\hspace{1mm}Hiroki Mori} \\
	Faculty of Engineering\\
	Utsunomiya University\\
	7-1-2 Yoto, Utsunomiya, 321-8585 Japan \\
	\texttt{hiroki@speech-lab.org} \\
}

\renewcommand{\shorttitle}{Spontaneously speaking conversational agent}

\begin{document}
\maketitle

\begin{abstract}
This study investigated the effect of synthetic voice of conversational agent trained with spontaneous speech on human interactants. Specifically, we hypothesized that humans will exhibit more social responses when interacting with conversational agent that has a synthetic voice built on spontaneous speech. Typically, speech synthesizers are built on a speech corpus where voice professionals read a set of written sentences. The synthesized speech is clear as if a newscaster were reading a news or a voice actor were playing an anime character. However, this is quite different from spontaneous speech we speak in everyday conversation. Recent advances in speech synthesis enabled us to build a speech synthesizer on a spontaneous speech corpus, and to obtain a near conversational synthesized speech with reasonable quality. By making use of these technology, we examined whether humans produce more social responses to a spontaneously speaking conversational agent. We conducted a large-scale conversation experiment with a conversational agent whose utterances were synthesized with the model trained either with spontaneous speech or read speech. The result showed that the subjects who interacted with the agent whose utterances were synthesized from spontaneous speech tended to show shorter response time and a larger number of backchannels. The result of a questionnaire showed that subjects who interacted with the agent whose utterances were synthesized from spontaneous speech tended to rate their conversation with the agent as closer to a human conversation. These results suggest that speech synthesis built on spontaneous speech is essential to realize a conversational agent as a social actor.
\end{abstract}


\section{Introduction}
\label{sec:introduction}

Nass and Moon\cite{Nass2000} stated that humans interacting computers perceive them as social actors, and unconsciously behave socially towards them. In their several socio-psychological experiments, they observed that humans react to computers in the same social manner as they do to humans. These findings, however, do not mean that humans always respond to computers as they do to humans. Rather, people rarely exhibit anthropocentric reactions to computers that surround us today.
They treat voice assistants\cite{Bentley2018,Hoy2018} as mere machines that can be controlled by voice commands. People rarely respond to voice agents with backchannels such as ``Uh-huh,'' interjections such as ``Wow!,'' or emotional expressions such as laughs, unlike they do to humans. To make human-computer interaction closer to human-human interaction, we think it necessary to change the current human attitude toward computers to be more social. In the ``Computers are social actors'' theory\cite{Nass1994}, cues that are closely associated with the human prototype, such as spoken words and interactivity, trigger prescribed behaviors in human-human interactions, which in turn cause mindless social responses for machines. This suggests the possibility that more human-like cues from machines encourage more social responses from humans. We are approaching the realization of such a conversational agent as a social actor from the viewpoint of speech synthesis. We assume that the way a conversational machine has \emph{social patiency} \cite{Jackson2021} (= the capacity to have its \emph{face}\cite{Brown1987} threatened or affirmed by social action) reflects its \emph{paralanguage} (= the way of speaking) as an \emph{observable} that constitutes the human interactant's \emph{levels of abstraction} (LoA) \cite{Floridi2008}. In other words, we assume that humans do not take social actions to a conversational machine as they do to a human partly because current speech synthesis lacks some paralinguistic property that constitutes their LoA at which the machine is perceived as having social patiency.

Nonverbal social behavior such as backchannels, laughs, expressive interjections, and repetitions is an indication of attending conversations, and helps a lot in facilitating human-human conversations, which Den and his colleagues\cite{Den2012} called \emph{response tokens}. Backchannels are short utterances spoken by a listener such as ``Uh-huh'' or ``Yeah''. Maynard defined it as a brief expression sent by the listener while the speaker is speaking\cite{Maynard1993}. The role of the backchannel of a listener is to display that she/he understands what the speaker is saying or is paying attention to the speaker \cite{Horiguchi1988}. Expressive interjections are non-lexical speech sounds which indicate the speaker's cognitive or affective state changes. They are distinguished from another type of interjections, filled pauses, in their unique morphological and pragmatic properties\cite{Mori2015}. Repetitions have social functions, such as involvement in conversations, showing interest, concerning and surprising, and eliciting backchannel-like responses \cite{Beun1995}. In addition to these response tokens, filled pauses play an important role in conversations. Studies on filled pauses suggest that they show hesitation or intention to take a turn\cite{Yamane2002,Mizukami2007}.

In a human-human conversation, the speaker takes advantage of these social responses as indicators of the listener's state of understanding and dynamically redesigns her or his speech accordingly. Such social responses could be useful for smoother interaction between human and computer as well, for example, regulating computer's speech rate and response timing, once humans begin to show social responses to computers.

Studies on spoken dialogue systems involve dialogue management\cite{Thomson2010,Xu2016}, speech recognition\cite{Graves2014,Chorowski2015,Chan2016}, and speech synthesis\cite{Tokuda2000, Zen2009, Zen2013, Tokuda2013,Wang2017,Shen2018,Ren2020}.
Recent studies on speech synthesis\cite{Wang2017,Shen2018,Ren2020} have dramatically improved the quality of synthesized speech to the point where it is difficult to distinguish the synthesized speech from a real human voice. 
However, this does not mean that current speech technology has reached a status where it can fully reproduce all speech sounds that humans may utter in everyday conversation.
So far, the construction of speech synthesizers has been assumed to involve written sentences read by voice professionals such as newscasters or voice actors. For this reason, current speech synthesizers speak as if a newscaster were reading a news or a voice actor were playing an anime character. In daily conversation, however, we speak spontaneously without a script. Spontaneous speech
has quite different nature from read speech, especially in its prosody \cite{Koriyama2011,Koriyama2012}. Speech in conversation may exhibit some characteristic tone patterns including the final rise-fall\cite{Pierrehumbert1988,Venditti1998,Ishi2006}. These prosodic cues are considered to be crucial not only to sound like spontaneous speech, but also play a meta-communicative role such as turn coordination, which is intrinsic to real-time interactions of humans. Using a read speech corpus on scripted text for building a speech synthesizer is to ignore such an important aspect of speech in conversation.

A number of studies aim to apply the nonverbal aspects of human-human interactions, such as turn-taking\cite{Skantze2017}, anthropomorphization\cite{Kontogiorgos2019}, backchannel\cite{Krogsager2014}, and gaze\cite{Skantze2013}, to enhance human-computer interactions.
Regarding nonverbal aspects of synthesized speech for spoken dialogue systems, Chiba et al.\cite{Chiba2018} showed the effects of emotional speech synthesis in non-task-oriented dialogue systems. The study showed that the use of appropriate emotional expression could improve subjective impressions of humans such as dialogue richness and likeness to the agent. James et al.\cite{James2018} showed that adding appropriate emotion to the speech of the robot could express empathy for users and be preferred to a typical voice of the robot. Misu et al.\cite{Misu2012} investigated whether the dialogue- or monologue-style induced more backchannels from humans as training data for speech synthesizers. They showed that synthesized speech in dialogue-style induced more natural backchannels and nods than in monologue-style.

In almost all previous studies on spoken dialogue system including the above, the speech synthesizer used was the one trained with read speech. Despite the discrepancy between read and spontaneous speech as pointed out above, there have been quite few attempts to synthesize conversational speech using a spontaneous speech corpus. After an attempt at HMM-based speech synthesis using conversational speech\cite{Andersson2010}, no work on spontaneous conversational speech synthesis was found in Interspeech conferences or SSW (Speech Synthesis Workshops) presentations until Ben-David and Shechtman\cite{Ben-David2021}, except for the authors' works \cite{Nagata2013, Yokoyama2018}. One reason for this might be that it has been unclear what a spontaneously speaking machine could be useful for. The current study is the first one to demonstrate that speech synthesis based on spontaneous speech is actually useful for making human-machine interaction more like human-human interaction.

Our study investigates the effect of synthetic voice of conversational agent trained with spontaneous speech on human interactants. We hypothesize that the synthetic voices of current conversational agents, built on a read speech dataset, cause humans to behave as if they are mere machines rather than \emph{social actors}.
We also hypothesize that humans will exhibit more social responses when interacting with a conversational agent that has a synthetic voice built on a spontaneous speech dataset.

To test these hypotheses, we focused on nonverbal behavior of humans as listeners, which include backchanneling, interjections, laughing, and nodding because humans rarely exhibit this behavior while interacting with the existing spoken dialogue systems. We also observed the response time of human interactants. Timing is important because, in human-human communication, people attend closely to the turn-taking timing of their own and their partner, trying to keep the right timing when it goes awry\cite{Clark2002}. On the other hand, if people do not regard a conversational agent as a social actor, they would not care about when to speak, and they would not mind keeping the agent waiting all the time. So far, the nonverbal behavior of spoken dialogue systems has been studied extensively \cite{Skantze2017,Kontogiorgos2019,Krogsager2014,Skantze2013}. However, studies regarding nonverbal behavior of users have been very few\cite{Misu2012,Bliek2020,Hjalmarsson2012,Arimoto2019}. The conversation experiment designed in this paper focuses on the nonverbal behavior of human interactants. This allows us to examine whether humans produce more nonverbal responses to a spontaneously speaking conversational agent in shorter delays, i.e., their tendency to behave as \emph{social agents}\cite{Jackson2021}.

For the experiment, we set up two conversational agents. One is an agent whose utterances were synthesized using spontaneous speech, and the other is an agent whose utterances were synthesized using read speech. Two groups of subjects, assigned to either the ``spontaneous'' or ``read'' condition, participated in a chat with the conversational agent that follows an identical scenario. Then, the frequency of the subjects' response tokens and the distribution of response time are compared for the two conditions. A subjective evaluation using a questionnaire on the impressions of the conversational agent was also conducted.

\section{Speech synthesis\label{synthesis}}
\subsection{Corpus}
\subsubsection{Spontaneous speech corpus}
For the spontaneous speech corpus, we used the Utsunomiya University Spoken Dialogue Database (UUDB) \cite{Mori2011}, where participants (12 females and 2 males) were engaged in the ``four-frame cartoon sorting task'' to estimate the original order of the shuffled frames. This corpus consists of 27 sessions and lasts about 130 min. In this study, utterances of a female speaker FTS were used because her recorded speech was longest in total duration in this corpus (about 18 min.).

\subsubsection{Read speech corpus}
For the read speech corpus, we used the Japanese speech corpus of Saruwatari-lab, the University of Tokyo (JSUT) \cite{Sonobe2017}. JSUT contains speech data of a female who is not a professional speaker but has experience working with voices. We used all the subcorpora of JSUT, which are approximately 10 hours in length in total.

\subsection{Method}
\label{sec:speechsynthesis}
As the speech synthesizer, we used Tacotron~2 \cite{Shen2018}. Tacotron~2 is a combination of the Tacotron-style model\cite{Wang2017} that generates mel-spectrograms from a sequence of characters and the modified WaveNet \cite{Giviansky2017,Tamamori2017} that conditions on mel-spectrograms to generate the waveform. Since this approach can directly learn the correspondence between characters and waveforms, it can generate speech of such high quality that it is difficult to distinguish from a real human voice.

In this study, we replaced the neural vocoder with MelGAN \cite{Kumar2019}. To train MelGAN, we used spontaneous monologue speech of 361 speakers in the Corpus of Spontaneous Japanese \cite{Maekawa2003}.

Two Tacotron~2 models were trained for synthesizing the agent's utterances to be used in the conversation experiment. The ``read'' model was built on the read speech corpus, JSUT. Likewise, the ``spontaneous'' model was built on the spontaneous speech corpus, UUDB. However, model training could not be done straightforwardly unlike the ``read'' model, because UUDB is a natural dialogue corpus. Treatment of nonverbal sounds such as laughter was an issue. For the present study, we simply ignore them; utterance that contains nonlinguistic sounds was split so as to include spoken contents only. Another unique phenomenon in spontaneous speech is filled pauses and expressive interjections. Because these interjections have different acoustic properties from ordinary lexical sounds \cite{Maekawa2017,Mori2015}, a dedicated vowel set was defined to transcribe these sounds for UUDB.

Another issue in using a natural dialogue corpus to build speech synthesizers is its insufficiency in size. To overcome this, the pretraining and fine-tuning approach was adopted. Starting from ``read'' model trained from JSUT with sufficient amount of data, the ``spontaneous'' model was trained by fine-tuning the initial model using UUDB. This allowed us to obtain a near conversational synthesized speech with reasonable quality, even with a small amount of data.

In this study, the mel spectrograms were calculated using a short-time Fourier transform using 50~ms frame size, 12.5~ms frame hop, and a Hann window function, as in the original Tacotron~2 paper \cite{Shen2018}. The model hyperparameters were set to the default values of the NVIDIA's implementation \cite{Nvidia}, except for the threshold to generate the stop token to be 0.1, which was necessary to obtain stable outputs.

\subsection{Analysis of prosody}
\label{sec:prosody}
The ``spontaneous'' model produces speech that gives a quite different impression than conventional speech synthesizers built on read speech corpora, primarily due to its prosody. Specifically, the ``spontaneous'' model reproduces tone patterns characteristic of conversational speech, particularly phrase-final tones.
To quantitatively compare the prosody synthesized by the ``spontaneous'' and ``read'' models, a prosodic labelling based on the J\_ToBI \cite{Venditti2005} (Japanese Tones and Break Indices) was performed for the synthesized speech. J\_ToBI describes prosody from two aspects; prosodic pitch (Tone) and prosodic boundary strength (BI) in Japanese. It defines a set of phrase-final boundary tone labels L\%, L\%H\%, and L\%HL\%, which roughly correspond to fall, rise, and rise-fall, and the latter two combined tones constitute the boundary pitch movements (BPMs). Because most of the sentences in read speech corpora are declarative, and voice professionals generally avoid using BPMs except at the end of a sentence when reading texts out loud, conventional speech synthesizers are exclusively capable of producing speech without BPMs, unless interrogative sentences are specially handled. Contrastively, spontaneous speech contains a lot of BPMs. Therefore, we assumed that the ``spontaneous'' model produces speech with a larger number of BPMs than the ``read'' model for a fixed set of text.

\begin{table}[tb]
  \caption{Frequencies of the phrase-final boundary tones appeared in the conversational agent's synthesized utterances.}
  \label{tab:jtobi}
  \centering
  \begin{tabular}{|l||c|c|}\hline
    & ``read'' model& ``spontaneous'' model\\ \hline
    L\% & 133 & 105 \\ \hline
    L\%H\% & 9 & 20 \\ \hline
    L\%HL\% & 0 & 17 \\ \hline
 \end{tabular}
\end{table}

The distribution of phrase-final boundary tones in the synthesized utterances using the two models is shown in Table~\ref{tab:jtobi}. Note that since the set of input texts was exactly the same, the total number of phrase-final boundary tones (corresponding to Break Index 2 \cite{Venditti2005}) was also the same for the two models.
This result shows that there was no rise-fall pattern in the utterances synthesized with the ``read'' model, in contrast to the ``spontaneous'' model. This means that utterances synthesized with the ``spontaneous'' model reflected the prosodic properties of natural conversational speech used to train the model.

\subsection{Subjective evaluation of synthesized speech}

To compare the overall impressions of synthesized speech built from ``read'' and ``spontaneous'' models, a subjective evaluation test was performed. In addition to speech \emph{clarity} as a common criterion in evaluating speech synthesizers, we also evaluated speech \emph{spontaneity}, the degree to which the synthesized speech sounds like it was uttered on the spot without a script. The Likert scales for assessing clarity and spontaneity in the questionnaire were:
\subsubsection*{Clarity}
\begin{etaremune}
\item Excellent
\item Good
\item Fair
\item Poor
\item Bad
\end{etaremune}
\subsubsection*{Spontaneity}
\begin{etaremune}
\item I am convinced that she was speaking what came to her mind on the spot.
\item I feel that she was speaking what came to her mind on the spot.
\item I am not sure whether she was speaking what came to her mind on the spot or a script.
\item I feel that she was speaking from a script.
\item I am convinced that she was speaking from a script.
\end{etaremune}

The subjective evaluation test was conducted as a follow-up to the conversation experiment described in Sect.\ \ref{sec:experiment}.
The subjects were 26 undergraduate and graduate students who also participated in the conversation experiment and agreed to participate the additional experiment.
The stimulus set consisted of 50 synthesized utterances, which was identical to that used in the conversation experiment described in the next section, per model. The experiment was conducted in a within-subjects design, i.e., each subject evaluated a total of \(50\times2=100\) utterances.

\begin{figure}[tb]
  \centering
  \includegraphics[width=.5\hsize]{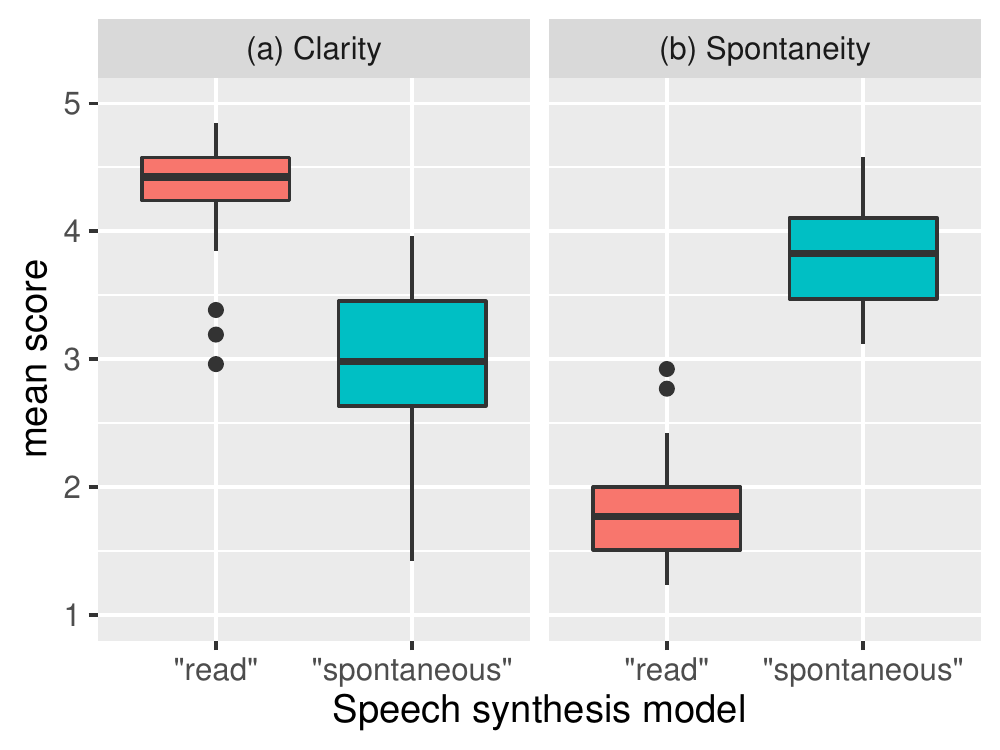}
  \caption{Comparison of synthesized speech with the ``read'' and ``spontaneous'' models from the viewpoints of clarity and spontaneity.}
  \label{mos}
\end{figure}

The results of the subjective evaluation test are shown in Fig.~\ref{mos}.
In the box-and-whiskers plot, the lower and upper hinges correspond to the first and third quartiles, the lower and upper whiskers extend from the hinge to the smallest and largest not-outlying values, and individual points correspond to the outlying values.
The mean clarity was 4.34 and 2.93 for the ``read'' and ``spontaneous'' models, respectively (Fig.~\ref{mos}(a)). From this result, it can be said that synthesized speech built on the read speech corpus was perceived clearer as a whole than that built on the spontaneous speech corpus.
Note, however, that this does not necessarily mean the inferiority of the speech synthesis based on spontaneous speech. Rather, we should argue whether the synthesized speech is intelligible enough for smooth communication. It is natural to assume that our daily speech is inherently less clear than read-aloud speech in, for example, news or dramas, since we try to save as much speech effort as possible to the extent that the speech act can achieve its goal. Conversely, one would feel unnatural if someone else spoke to her/him as clearly as a newscaster reads a news. From this perspective we think that the clarity of the ``spontaneous'' model is acceptable for the conversational agent's voice.

The mean spontaneity was 1.80 and 3.78 for the ``read'' and ``spontaneous'' models, respectively (Fig.~\ref{mos}(b)). This result indicates that the ``spontaneous'' model tends to produce speech that sounds as if it was uttered on the spot more than the ``read'' model.

From these results, it was found that the speech synthesized with the ``spontaneous'' model indeed had a quality of spontaneous speech. By using a spontaneous speech corpus for training, it is possible to synthesize speech that is close to our everyday speech, at least in some aspects. Therefore, one might expect a human-machine interaction that is closer to a human-human interaction, by using ``spontaneous'' synthesized speech as the machine's voice.

\section{Conversation experiment with a spontaneously speaking agent}
\label{sec:experiment}
\subsection{Overview of the conversational agent}

\begin{figure}
  \centering
  \includegraphics[width=.5\hsize]{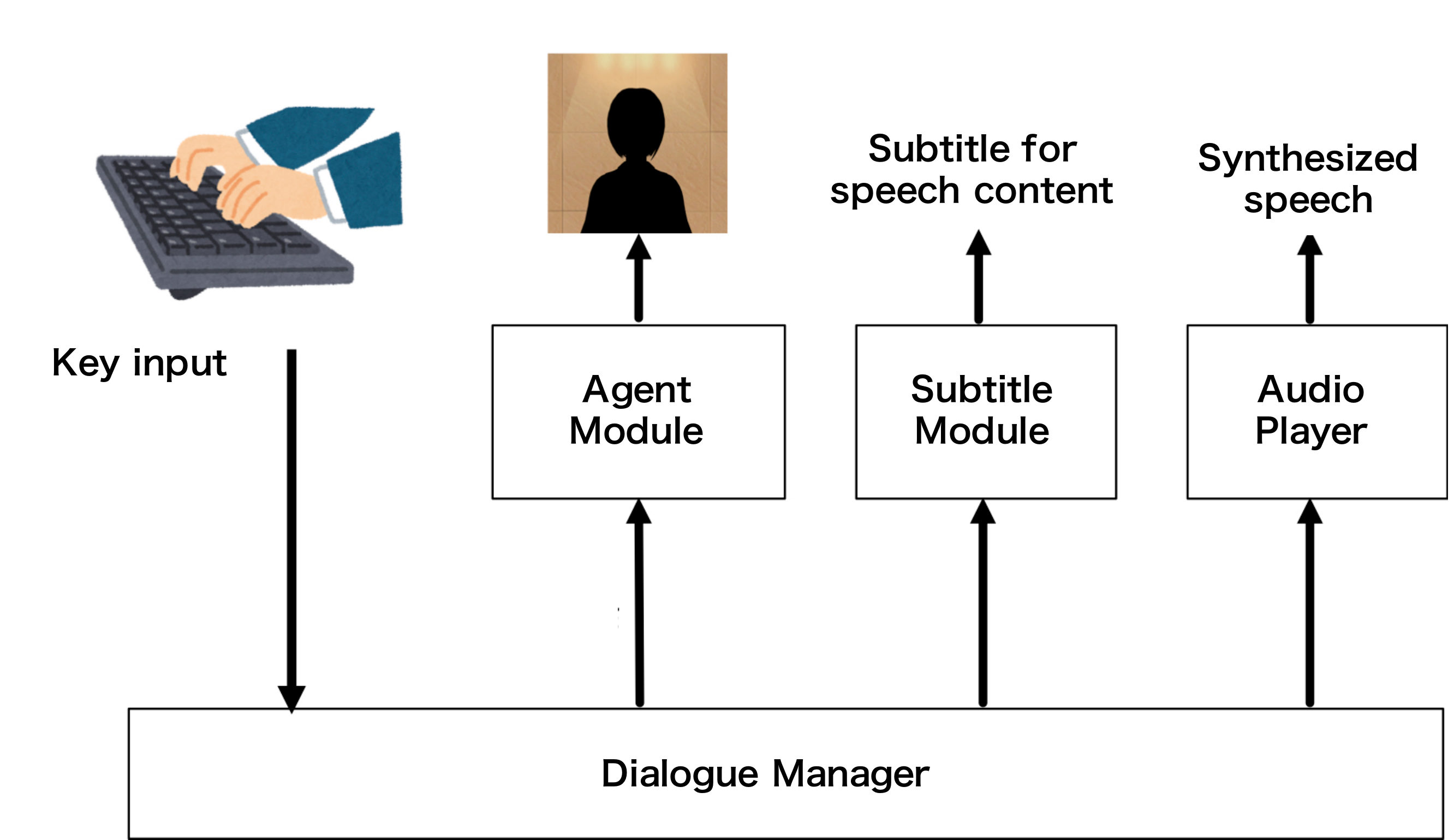}
  \caption{Overview of the conversational agent (adapted from \cite{Lee2013}.)\label{mmdagent}}
\end{figure}

The conversational agent used for this experiment was implemented using the MMDAgent \cite{Lee2013}. MMDAgent is a platform for building spoken dialogue systems that have modules of speech recognition, speech synthesis, dialogue management, and 3D model motion management. Fig.~\ref{mmdagent} shows an overview of the conversational agent used in this experiment. 
In this study, we did not use the default speech recognizer but instead applied the Wizard of Oz (WoZ) technique \cite{Fraser1991}, where an experimenter operates the agent behind the subject. The reason for employing WoZ was to prevent bad impressions to the agent due to speech recognition errors or unnatural speech timings. The agent was designed to speak by playing back pre-synthesized utterances according to the wizard's operation.

The dialogue scenario was designed so that the agent speaks almost unilaterally. In the scenario, the agent asks the human interactant if she/he is interested in traveling abroad, right after the initial greeting. No matter she/he is interested or not, the agent talks about countries she loves to go. After this, the agent continues to talk about various trivia about countries around the world. Sometimes she quizzes the human interactant, such as ``Do you know which country is most famous for pyramids?'' Basically, answering to these quizzes are only opportunity for the human interactant to take turns, forcing she/he to be a listener for the rest of the time.

Instead of writing the scenario by hand, we first recorded a dialogue between one of the authors and his close relative, where the author improvised the role of the agent, with no script at all. The recorded dialogue was then transcribed and transformed to an FST (finite-state transducer) for \mbox{MMDAgent} by an in-house tool. We consider it crucial to avoid handwritten scripts for the utterances of conversational agents, because actual words of spontaneous utterances have different linguistic characteristics from ``imaginary'' words, and human interlocutors tend to behave differently in response \cite{Takamatsuya2020}.

The wizard manipulated the agent's behaviors, which include triggering the next utterance in the scenario, determining whether the subject's answer to a quiz is correct or not, triggering a backchannel, triggering an utterance to encourage the subject to speak friendly, triggering a confirmation that the subject is attending, and triggering an expression to get the conversation back on track (such as ``Anyway,'') when it is going to break down. Determining the correctness of the answers to the quiz was necessary to reflect on the agent's next action. Sending backchannel was intended to make the agent behave more human-like when the subject is speaking. A previous study revealed that randomly generating acoustically different backchannels improves the naturalness of dialogue compared to repeatedly generating an identical backchannel \cite{Mori2013}. Therefore, three similar but different backchannels were prepared and randomly selected for playback. The purpose of encouraging to speak friendly was to induce a relaxed and natural mood, as if the subject were talking with a friend. The purpose of asking if the subject was listening was to prevent the subject from becoming a mere listener and to encourage reactions. However, this operation was limited to twice at most in each conversation.

The appearance of the agent was replaced to a silhouette in order to prevent any inconsistency between the appearance of the agent and the individuality of the synthesized speech.

The system displayed subtitles simultaneously with the agent's utterance. This prevented subjects from missing utterances even when the quality of synthesized speech was not sufficient.

\subsection{Method}
The subjects were 50 undergraduate and graduate students who were not engaged in speech research. They received both verbal and written explanations of the experiment, and provided written informed consent before the experiment. The experiment was approved by the Ethics Committee on Research Involving Humans, Utsunomiya University.

They were assigned to either the ``spontaneous'' or ``read'' condition, namely, the experiment was conducted in a between-subjects design. In the ``spontaneous'' condition, the subject had a conversation with the agent whose utterances were synthesized by the ``spontaneous'' model described in Sect.~\ref{sec:speechsynthesis}. The ``read'' condition was identical to the ``spontaneous'' condition except that the agent's utterances were synthesized by the ``read'' model. A video excerpt of a conversation in the ``spontaneous'' condition is included as supplemental material.

It is not considered fair to have a subject participate in both conditions. If a subject interacted with both ``spontaneous'' and ``read'' agents, she/he would easily notice that the objective of the experiment was to test the effect of the agent's voice. Eventually, the subject might also notice that one agent was speaking spontaneously unlike existing dialogue systems and, in an effort to be a ``good subject,'' might try to behave more favorably in her/his interaction with the ``proposed'' agent. Therefore, the conversation experiment should not be conducted in a within-subjects design, but in a between-subjects design.

As evaluation indices of how close the human-agent interaction and human-human interaction are, we examined the response time of subjects to the agent, the number of subjects' response tokens (backchannels, expressive interjections, laughs, and filled pauses), and the number of nods.
By investigating these nonverbal behaviors, it is possible to determine whether humans talking with a conversational agent behave as if it were a human-like social actor and not just a machine.

After the session was over, each subject was asked to rate her/his impressions of the agent and the quality of the conversation using a 6-item questionnaire. As a debriefing after the evaluation, the subjects were told that the conversational agent was not automated but human-operated.

\section{Result}

\subsection{Nonverbal behavior}
\begin{figure}[tb]
  \centering
  \includegraphics[width=.7\hsize]{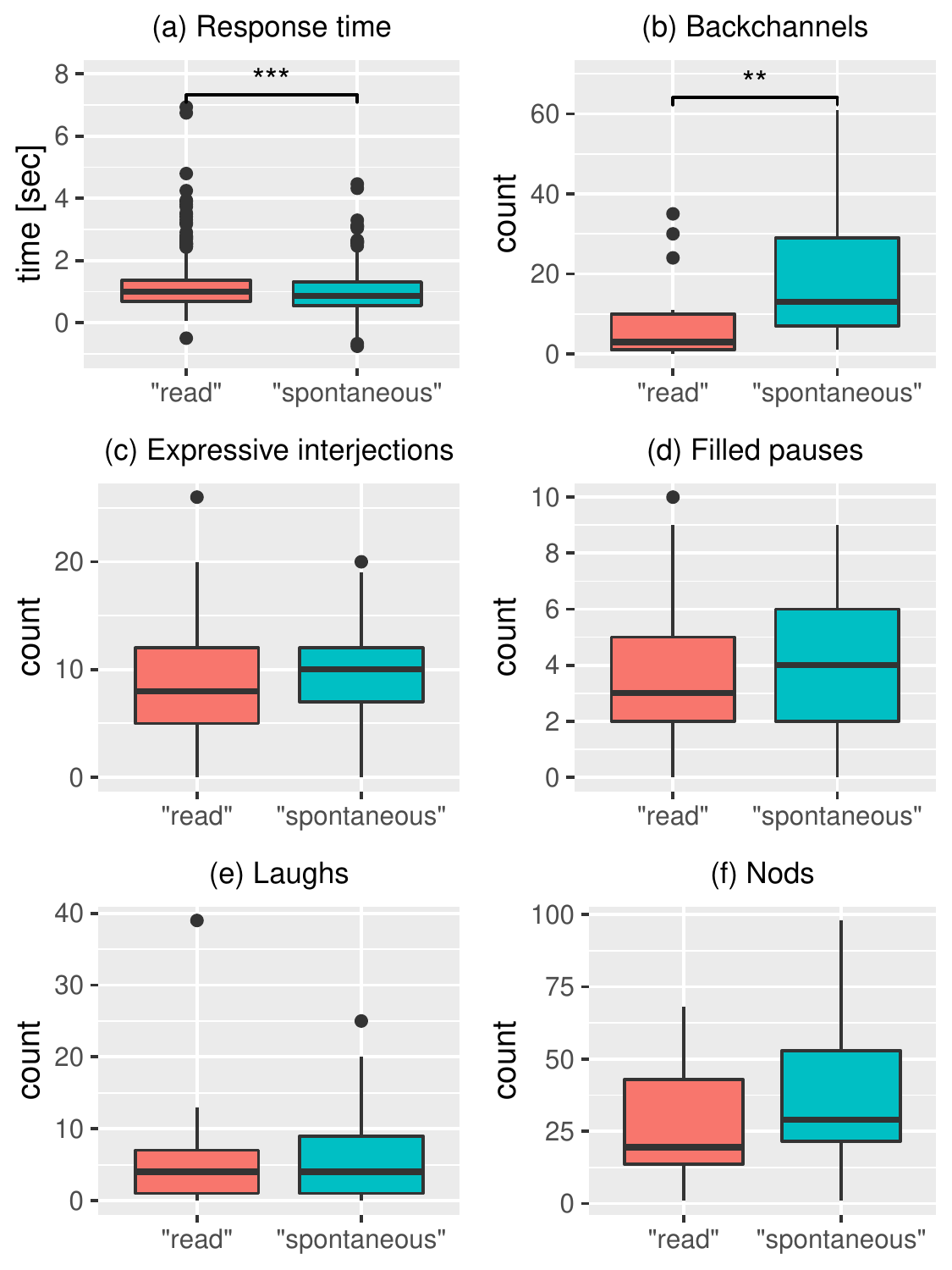}
  \caption{Nonverbal behavior indices.}
  \label{subject_behavior}
\end{figure}

Fig.~\ref{subject_behavior} shows the distribution of the indices for nonverbal behaviors during the interaction between the conversational agent and subjects.
In the following paragraphs, summary statistics are shown in the form of mean and 95\% confidence intervals.

The mean response time was \(1.16\pm0.07\) sec.\ for the ``read'' condition and \(0.99\pm0.05\) sec.\ for the ``spontaneous'' condition (Fig.~\ref{subject_behavior}(a)), and the difference was significant (Welch's $t$-test, \(t(957.8)=3.80\), $p=0.00016$). This result indicates that subjects who interacted with the agent whose utterances were synthesized from spontaneous speech data tended to respond faster than those who interacted with the agent whose speech was synthesized from read speech data.

The average number of backchannels was \(6.96\pm3.92\) for the ``read'' condition and \(19.00\pm6.91\) for the ``spontaneous'' condition (Fig.~\ref{subject_behavior}(b)), and the difference was significant (\(t(38.0)=-3.13\), $p=0.0034$). This result indicates that subjects who interacted with the agent whose utterances were synthesized from spontaneous speech data tended to show a larger number of backchannels than those who interacted with the agent whose speech was synthesized from read speech data.

The average number of expressive interjections was \(8.92\pm2.68\) for the ``read'' condition and \(10.12\pm2.02\) for the ``spontaneous'' condition (Fig.~\ref{subject_behavior}(c)), and the difference was not significant (\(t(44.6)=-0.71\), $p=0.48$).

The average number of filled pauses was \(3.60\pm1.09\) for the ``read'' condition and \(3.92\pm1.01\) for the ``spontaneous'' condition (Fig.~\ref{subject_behavior}(d)), and the difference was not significant (\(t(47.8)=-0.45\), $p=0.66$).

The average number of laughs was \(5.72\pm3.26\) for the ``read'' condition and \(6.04\pm2.82\) for the ``spontaneous'' condition (Fig.~\ref{subject_behavior}(e)), and the difference was not significant (\(t(47.0)=-0.15\), $p=0.88$).

The average number of nods was \(27.58\pm8.23\) for the ``read'' condition and \(36.42\pm10.20\) for the ``spontaneous'' condition (Fig.~\ref{subject_behavior}(f)), and the difference was not significant (\(t(44.1)=-1.40\), $p=0.17$).

In summary, humans interacting with the agent whose utterances were synthesized from spontaneous speech data tended to exhibit shorter response times and more response tokens, which can be interpreted that they behaved more like interacting with a human.

\subsection{Questionnaire}

Table~\ref{subjective_dialogue} summarizes the subjects' impressions of the agent and the quality of their conversations. Each section of the table shows the question, the meaning of the scale (5--1), and a contingency table (columns correspond to the response options and rows correspond to the conditions to which the subjects were assigned). The most notable result is for the question ``How close was your conversation with Mei-chan to a conversation with a human?''. The mean was 3.60 for the ``read'' condition and 4.08 for the ``spontaneous'' condition. The Brunner-Munzel test showed that the difference in the response distributions for the two conditions was significant ($p=0.049$). This result indicates that subjects who interacted with the agent whose utterances were synthesized from spontaneous speech data tended to evaluate their conversation as closer to a human conversation.
\begin{table}
  \centering
  \caption{Questions and responses asking subjects' impression of the agent and the conversation quality.}
  \label{subjective_dialogue}
  \includegraphics[width=.6\hsize]{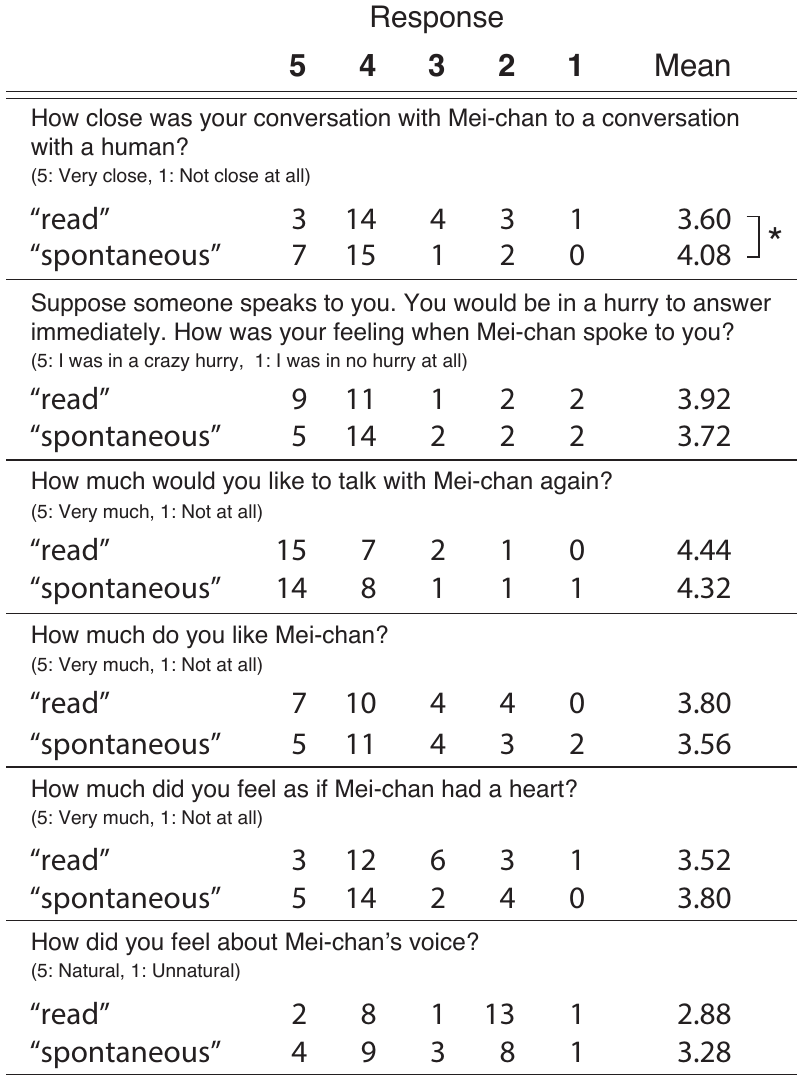}
\end{table}

The differences for other items were not significant.

\section{Discussion}

Subjects in the ``spontaneous'' condition tended to respond faster than in the ``read'' condition. This result suggests that subjects who interacted with the spontaneously speaking conversational agent are more likely to feel pressure to respond at the right timing. Since one is less likely to feel such time pressure when interacting with mere a machine, this implies that the subjects tended to view the agent as a social actor rather than mere a machine.

Subjects in the ``spontaneous'' condition tended to show a larger number of backchannels than in the ``read'' condition. Since one is less likely to show backchannels to mere a machine, this may also be evidence that the subjects tended to view the agent as a social actor rather than mere a machine.


Subjects in the ``spontaneous'' condition tended to evaluate their conversation with the agent as closer to a human conversation than in the ``read'' condition. This result supports the above interpretation of the nonverbal behavior results, i.e., spontaneously speaking agent tend to be more viewed as a social actor.

These results suggest that speech synthesis built on spontaneous speech is essential to realize a conversational agent as a social actor.

At this time, however, the specific features of spontaneous speech that explain these results still remain unknown.
One candidate for such a feature is the phrase-final boundary tones. The analysis described in Sect.~\ref{sec:prosody} showed a clear difference between utterances synthesized with the ``spontaneous'' model and those synthesized with the ``read'' model in terms of phrase-final boundary tones. Considering that the final rise-fall tones are characteristic of conversational speech \cite{Pierrehumbert1988,Venditti1998,Ishi2006}, and that they constitute important cues that determine the occurrence of backchannels in a computational model of dialogue prosody \cite{Koiso1998}, it is natural to attribute the occurrence of backchannel in the ``spontaneous'' condition to the L\%HL\% (rise-fall) pattern. To prove this, we tried to find the relationship between the rise-fall patterns of the agent's utterances and the subjects' subsequent responses, but could not find any direct relationship. Additional experiments will be needed, such as controlling the phrase-final boundary tones of the agent's utterances and seeing the effect on human behavior.

Another issue is the speaker individuality. If we had a pair of read speech corpus and spontaneous speech corpus of a same speaker, we could eliminate the extraneous variable, but building such a dataset would be very costly. We think the effect of speaker on the current experiment was minimal because the JSUT speaker and the UUDB speaker were both female and of the same generation.

This research is the antithesis of the conventional speech synthesis that has placed supreme importance on naturalness as professional speech. The current study suggests that speech synthesis for conversational agents should also aim to produce speech that sounds as if it were uttered on the spot. In the future, conversational agents will be used on a daily basis and will increasingly be treated as a partner or a friend, rather than just a tool. Speech synthesizers built on spontaneous speech will help to realize such a conversational agent. The current study clarifies the significance of using spontaneous speech for speech synthesis in the field of human-machine interaction research.

\section{Conclusions}
In this paper, we investigated the effect of synthetic voice of conversational agent trained with spontaneous speech on humans who interact with it. 
To quantitatively compare the prosody synthesized by the model trained with spontaneous speech and the model trained with read speech, a prosodic labeling was performed for the synthesized speech, and revealed that utterances synthesized with the model trained with spontaneous speech reflected the prosodic properties of natural conversational speech. A subjective evaluation test was also performed to assess the clarity and spontaneity of the synthesized speech, and the result showed that the model trained with spontaneous speech tended to produce speech that sounds more like spontaneous speech uttered on the spot. A large-scale conversation experiment was conducted with a conversational agent whose utterances were synthesized with either the model trained on spontaneous speech or that trained on read speech. This revealed that subjects who interacted with the agent whose utterances were synthesized from spontaneous speech tended to show shorter response times and a larger number of backchannels. Furthermore, the subjects who interacted with the agent whose utterances were synthesized from spontaneous speech tended to rate their conversation with the agent as closer to a human conversation.

In summary, it can be concluded that humans exhibit more social responses when interacting with a conversational agent that has a synthetic voice built on spontaneous speech, and such an agent is more likely to be viewed as a social actor.

\bibliography{mybib.bib}   
\bibliographystyle{hieeetr}

\end{document}